\documentclass{article}
\usepackage[T1]{fontenc}
\usepackage[utf8]{inputenc}
\usepackage[final]{ismir2026}
\usepackage{amsmath,cite,url}
\usepackage{amssymb}
\usepackage{graphicx}
\usepackage{color}
\usepackage{subcaption}  
\usepackage{float}
\usepackage{booktabs}

\title{Genre Bias or Aesthetic Perception? Identifying and Mitigating Shortcut Learning in Music Evaluation}

\multauthor
{
Yizhou Zhang$^{1\ast}$ \quad
Wangjin Zhou$^{1\ast}$ \quad
Yi Zhao$^{2}$ \quad
Wei Tan$^{2}$ \quad
Keisuke Imoto$^{1}$ \quad
Zhi Gong$^{2}$
}
{
$^{1}$ Graduate School of Informatics, Kyoto University, Japan\\
$^{2}$ WXG, Tencent, China\\
\small $^\ast$ Equal contribution
}
\def\authorname{Y. Zhang, W. Zhou, Y. Zhao, W. Tan, K. Imoto, and Z. Gong}

\begin{document}

\maketitle

\begin{abstract}
Music aesthetics scoring plays a critical role in applications such as dataset curation, generative model evaluation, and reward modeling for music generation. Recent approaches rely on deep neural networks trained on human-annotated ratings, but these models may exploit spurious correlations rather than capturing perceptually meaningful aesthetics. In this work, we identify a previously underexplored failure mode in music evaluation models: genre-induced shortcut learning. Through a systematic analysis of SongEval, we show that biases in training data lead to strong correlations between genre-related features and predicted scores, causing the model to use them as a proxy for aesthetics. This results in systematic overestimation of pop music and undervaluation of high-quality samples from other genres, leading to predictions that are inconsistent with human preferences. To address this issue, we propose a training objective that jointly reweights hard samples and regularizes group-level performance, encouraging the model to learn genre-invariant representations of musicality. Experimental results demonstrate that our method reduces genre-dependent bias and improves alignment with human preferences, as reflected by gains in both cross-genre and within-genre preference alignment.
\end{abstract}

\section{Introduction}\label{sec:introduction}

Automatic music aesthetics assessment plays an increasingly important role in modern music generation and evaluation pipelines \cite{lerch2025survey}. Music aesthetics scores are widely used for filtering and curating datasets, benchmarking music generation models \cite{grotschla2025benchmarking}, and providing reward signals to align generative systems with human preferences \cite{cideron2024musicrl}. Recent approaches typically rely on deep neural networks trained on human-annotated ratings to approximate subjective aesthetic judgments \cite{tjandra2025meta, yao2025songeval, liu2025musiceval}.

Although various objective metrics \cite{roblek2019fr, huang2022mulan, gui2024adapting} have been proposed to evaluate generated music, they often fail to capture the inherently subjective and multidimensional nature of musicality \cite{yuan2025yue}. Moreover, these metrics typically show weak correlation with human judgments of musical quality \cite{zhang2025aesthetics}. As a result, learning-based approaches have become increasingly dominant \cite{tjandra2025meta, yao2025songeval, liu2025musiceval}. Despite promising progress \cite{ma2026icassp}, a fundamental concern remains: these models may rely on spurious correlations in the training data rather than capturing perceptually meaningful quality cues.

In particular, human ratings collected from synthetic music corpora, often generated by diverse music generation models, exhibit systematic biases across sub-populations such as musical genres. Models trained on such data may therefore exploit these correlations as shortcuts to minimize training loss, leading to unreliable and potentially misaligned evaluation signals \cite{wei2026robust}.

In this work, we identify a critical and previously underexplored failure mode in music aesthetics evaluation: \textit{genre-induced shortcut learning}. Through a systematic analysis of SongEval \cite{yao2025songeval}, a widely used and reproducible reward model for music aesthetics, we show that biases in the training data induce strong correlations between genre-related features and predicted scores. Consequently, the model tends to rely on genre-related proxy for perceived aesthetics, resulting in a pop-centric evaluation behavior. This leads to systematic overestimation of pop music, undervaluation of high-quality recordings from other genres, and inconsistent cross-genre comparisons that deviate from human preferences. When used as reward models, such evaluators may encourage generative systems to imitate high-scoring genres rather than improve musical quality, reducing diversity and robustness.

To address this issue, we propose training strategies that explicitly counteract shortcut-driven optimization. Our approach encourages the learning of genre-invariant representations of musicality by reducing reliance on genre-related shortcuts while preserving perceptually relevant quality cues.

Experimental results show that our method reduces genre-dependent bias and improves alignment with human preferences, particularly in cross-genre comparisons, while also improving within-genre preference prediction.

Our contributions are summarized as follows:

\begin{itemize}
    \item We identify genre-induced shortcut learning in music aesthetics models, where biased and limited data lead models to exploit spurious correlations as proxies for perceived musicality.
    
    \item Through controlled experiments, human preference comparisons, and futher direction analysis, we demonstrate that this behavior results in systematic bias, including overestimation of pop music and undervaluation of high-quality recordings from lower-rated genres.
    
    \item We propose regularization strategies that encourage genre-invariant representations, mitigating bias while maintaining and improving alignment with human judgments.
\end{itemize}

\section{Problem Setup}
\label{sec:problem}

Let $z=(z_q,z_g,z_r)$ denote the representation of a music sample, where
$z_q$ captures quality-relevant factors, $z_g$ encodes genre identity, and
$z_r$ denotes other residual variation.

Ideally, aesthetic judgments should depend on genre only through
quality-relevant musical factors. We therefore assume that, under the
population distribution $p$,
\begin{equation}
y \perp (z_g, z_r) \mid z_q,
\end{equation}
or equivalently,
\begin{equation}
p(y \mid z)=p(y \mid z_q).
\end{equation}
Under the squared loss, the population-optimal predictor is the Bayes
predictor
\begin{equation}
f_p^*(z)
=
\operatorname*{arg\,min}_{f}
\mathbb{E}_{p}\big[(y-f(z))^2\big]
=
\mathbb{E}_{p}[y \mid z]
=
\mathbb{E}_{p}[y \mid z_q],
\end{equation}
where the last equality follows from the conditional independence assumption.

In practice, however, training data are sampled from an empirical distribution
$\hat p$, where genre imbalance and genre-dependent annotation biases may
induce a conditional dependence between labels and genre:
\begin{equation}
\hat p(\hat y \mid \hat z_q, \hat z_g)
\neq
\hat p(\hat y \mid \hat z_q),
\end{equation}
or equivalently,
\begin{equation}
\hat y \not\perp \hat z_g \mid \hat z_q
\quad \text{under } \hat p .
\end{equation}
Under squared-loss training, the empirical risk minimizer is therefore
\begin{equation}
f_{\hat p}^{*}(\hat z)
=
\operatorname*{arg\,min}_{\hat f}
\mathbb{E}_{\hat p}\big[(\hat y-\hat{f}(\hat z))^2\big]
=
\mathbb{E}_{\hat p}[\hat y \mid \hat z],
\end{equation}
which may exploit $\hat z_g$ as a predictive shortcut.

We quantify this shortcut dependence by the conditional mutual information
\begin{equation}
I_{\hat p}(\hat y; \hat z_g \mid \hat z_q).
\end{equation}
This quantity is zero when genre provides no additional information about the
label beyond quality-relevant factors, and becomes positive when the empirical
data make genre predictive of the aesthetic score. Other residual factors may
induce similar shortcuts, but this work focuses on genre-induced shortcut
learning.

\section{Diagnostic Analysis}
\label{sec:diagnostic}
We conduct a series of diagnostic analyses on SongEval~\cite{yao2025songeval}, a widely used music aesthetics evaluation model~\cite{chen2025diffrhythm+, gong2025ace, leilevo, liu2025jam, li2026songecho}. Accordingly, we analyze genre-related shortcut learning through a progressive diagnostic pipeline: 
first verifying spurious genre--score correlations in the training data; 
then testing whether these shortcuts transfer to model predictions under mildly controlled, genre-balanced evaluation; 
further examining their impact on human-preference alignment in same-genre and cross-genre comparisons; 
and finally analyzing the association between predicted aesthetic scores and pop-likeness.

\subsection{Diagnostic Setup}

SongEval provides five aesthetic evaluation dimensions. We first examine their pairwise correlations and observe that all correlations exceed $0.95$, indicating strong redundancy across dimensions. We therefore use the averaged score as a unified aesthetic measure.

To probe genre-dependent behavior, we construct two genre-balanced evaluation sets from real-world recordings: a subset of MTG-Jamendo~\cite{bogdanov2019mtg}
with 10 genres (50 samples per genre), and a subset of M6~\cite{li2026m6} with 10 genres (100 samples per genre). These datasets contain diverse musical content across genres and provide a balanced evaluation setting.

In addition, we incorporate preference-based evaluation using CMI~\cite{ma2026cmi}, including both human-annotated (CMI-Pref) and pseudo-labeled (CMI-Pref-Pseudo) pairwise comparisons. To ensure sufficient statistical reliability and coverage, we focus on the five most frequent genres in the dataset.

For CMI, we obtain genre labels using Qwen3-Omni \cite{xu2025qwen3}, a multimodal model capable of capturing high-level audio semantics. On MTG-Jamendo and M6, its predictions achieve over 90\% agreement with original annotations, indicating that it provides a reliable proxy for genre-related factors.

\subsection{Pop Dominated Distribution}
\label{subsec:pop}
\begin{table*}[t]
\centering
\caption{Genre distribution and average scores in SongEval training data.}
\label{tab:train_genre}
\begin{tabular}{lcccccccccc}
\toprule
Genre & pop & hiphop & classical & rock & country & electronic & blues & jazz & world & others \\
\midrule
Count & 1131 & 287 & 262 & 253 & 106 & 106 & 70 & 67 & 16 & 101 \\
Avg. Score & 3.50 & 2.90 & 2.73 & 3.17 & 3.43 & 3.03 & 2.80 & 2.44 & 3.23 & 2.65 \\
\bottomrule
\end{tabular}
\end{table*}

We first analyze the training distribution of SongEval. As shown in Table~\ref{tab:train_genre}, the dataset exhibits a highly imbalanced, long-tailed structure, where pop dominates the data, while most other genres are significantly underrepresented.

\begin{figure}
    \centering
    \includegraphics[width=\linewidth]{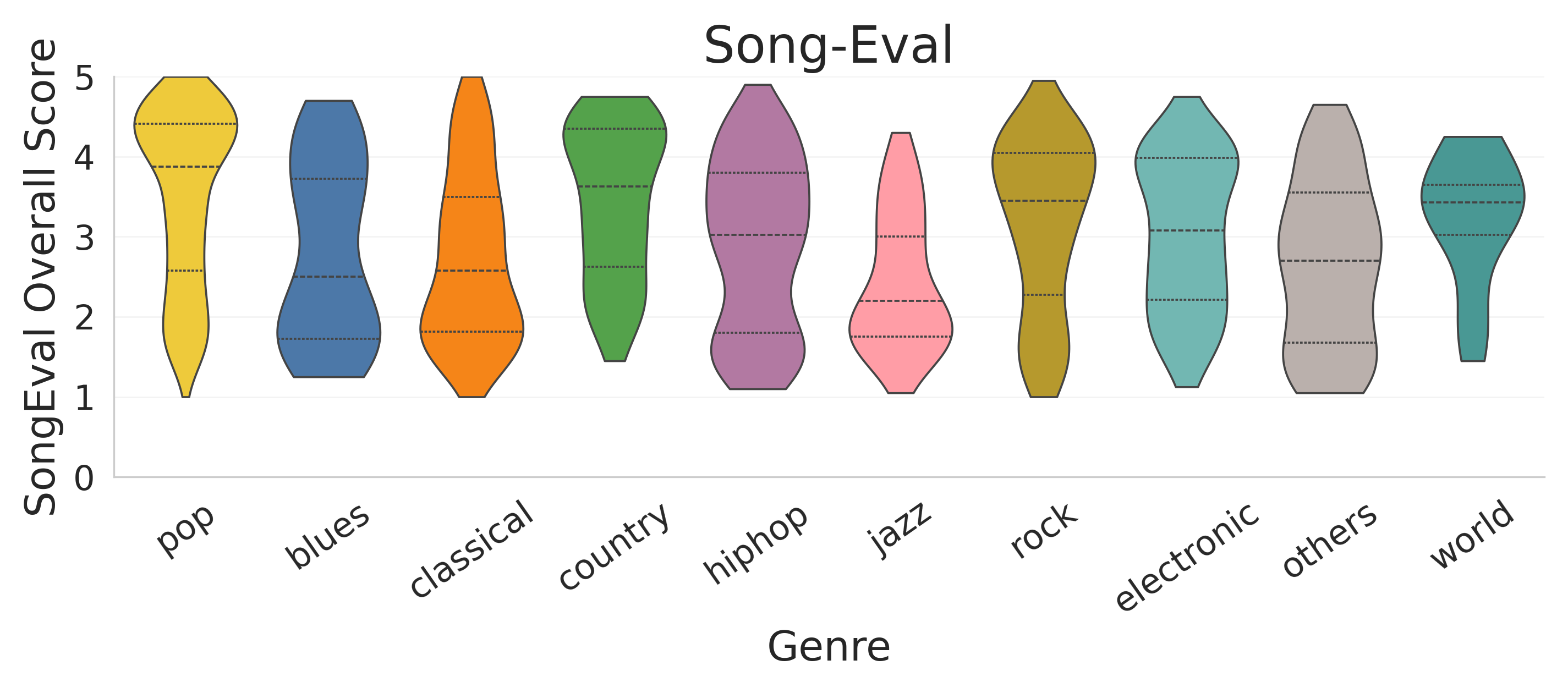}
    \caption{Score distributions across genres in SongEval training dataset}
    \label{fig:songeval}
\end{figure}

Beyond sample imbalance, the supervision signal is also skewed across genres revealed by Figure \ref{fig:songeval}. High-frequency mainstream genres tend to receive higher scores, whereas genres such as jazz and others are consistently assigned lower scores.

\subsection{Genre Bias under Mild Controlled Conditions}
\label{subsec:genre}
The design of MTG-Jamendo and M6 subset reduces two major confounders: genre imbalance and variation in production quality, thereby approximating a setting closer to 

\begin{equation}
    p(\hat y \mid \hat z) = p(\hat y \mid \hat z_q).
\end{equation}

Under this setting, if the learned predictor depends primarily on $\hat z_q$, its predictions should not exhibit systematic differences across genres.

However, we still observe significant genre-dependent variation in predicted scores. A one-way ANOVA test shows a strong genre effect ($p < 0.001$). The differences are substantial: for example, on MTG-Jamendo, pop achieves an average score around 3.5 while jazz remains below 2.5, and similar gaps are observed on M6. 

In both datasets, we observe a highly consistent pattern: pop and stylistically related genres receive higher scores, while genres such as jazz and classical receive lower scores. Despite differences in genre taxonomy and data sources, this pattern remains remarkably similar across MTG-Jamendo and M6, indicating that it is not dataset-specific.

Moreover, genres that are closer to pop in terms of production characteristics tend to receive higher scores, further indicating that the model relies on style-related cues as proxies for aesthetic quality.

\begin{figure}[t]
\centering
\includegraphics[width=\linewidth]{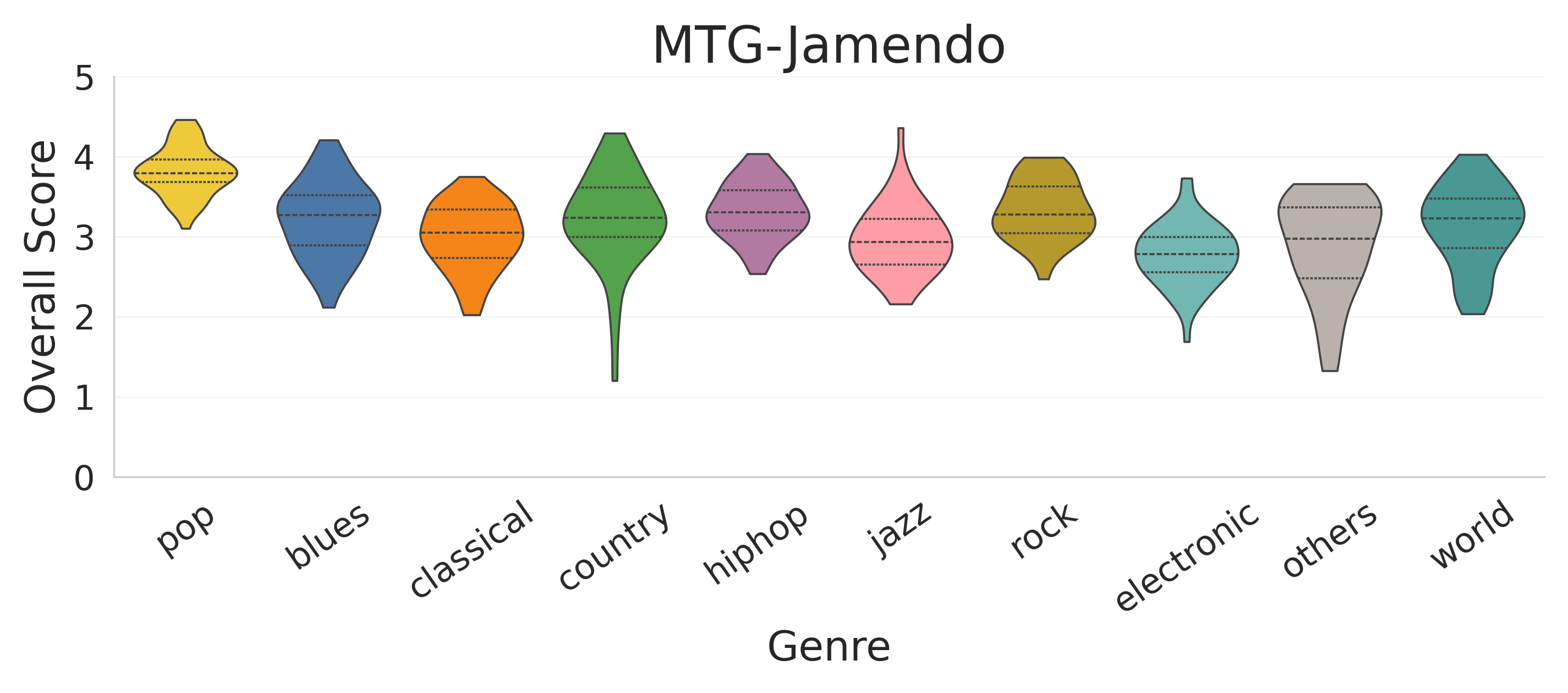}
\includegraphics[width=\linewidth]{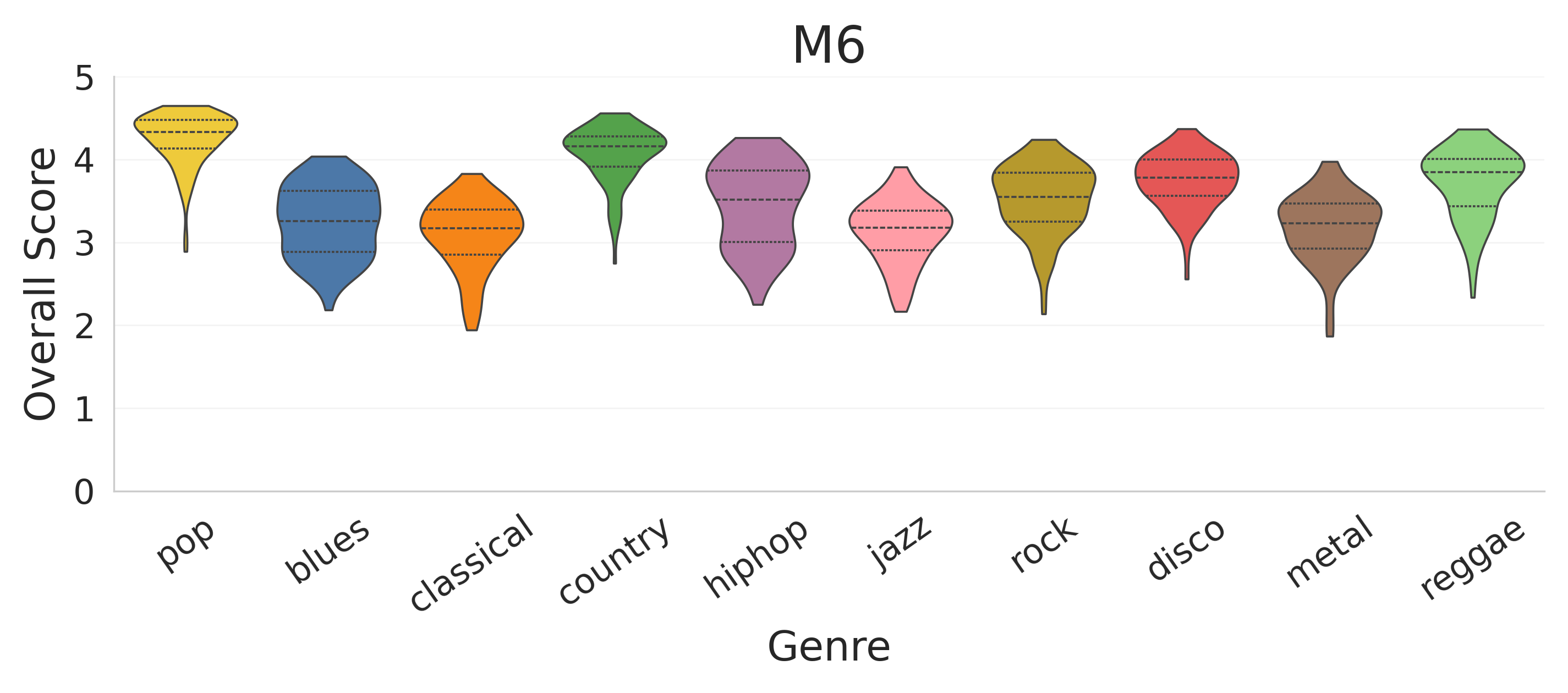}
\caption{Score distributions across genres in MTG-Jamendo and M6 datasets.}
\label{fig:genre_distribution}
\end{figure}

\subsection{Preference Discrepancy across Genres}
\label{subsec:preference}

While the controlled evaluation in Section~\ref{subsec:genre} removes major confounders, it does not directly assess alignment with human preferences. We therefore conduct a preference-based evaluation using pairwise comparisons from the CMI dataset. Overall, SongEval achieves an accuracy of 0.687 on CMI-Pref and 0.549 on CMI-Pref-Pseudo, indicating degraded performance under noisier preference signals.

Even within the same genre, performance varies substantially. On CMI-Pref, hiphop and rock achieve relatively higher accuracy (0.746 and 0.720), while classical, jazz, and pop are notably lower (from 0.64 to 0.66). This disparity becomes more pronounced on CMI-Pref-Pseudo, where jazz drops to 0.447 and classical to 0.544. Such variation under controlled genre conditions indicates that the model does not learn a stable evaluation criterion.

In cross-genre comparisons as shown in Table \ref{tab:cross_genre}, a clear directional pattern emerges: SongEval systematically overestimates pop while underestimating genres such as jazz and classical, with deviations that are large and asymmetric relative to human preference ratios. Moreover, these errors are not uniformly distributed, but concentrated on specific genre pairs, particularly those involving jazz and classical, which also exhibit weaker same-genre performance.

\begin{table*}[t]
\centering
\small
\setlength{\tabcolsep}{20pt}
\begin{tabular}{lccc|ccc}
\toprule
 & \multicolumn{3}{c}{CMI-Pref} & \multicolumn{3}{c}{CMI-Pref-Pseudo} \\
\cmidrule(lr){2-4} \cmidrule(lr){5-7}
Pair & Pred & GT & Acc. & Pred & GT & Acc. \\
\midrule
pop \textbf{v.s.} rock        & 0.774 & 0.645 & 0.587 & 0.591 & 0.245 & 0.576 \\
pop \textbf{v.s.} hiphop      & 0.824 & 0.832 & 0.795 & 0.754 & 0.594 & 0.733 \\
pop \textbf{v.s.} jazz        & 0.615 & 0.462 & 0.879 & 0.795 & 0.249 & 0.394 \\
pop \textbf{v.s.} classical   & 0.807 & 0.675 & 0.709 & 0.795 & 0.412 & 0.513 \\
rock \textbf{v.s.} hiphop     & 0.643 & 0.893 & 0.695 & 0.494 & 0.660 & 0.595 \\
rock \textbf{v.s.} jazz       & 0.000 & 0.250 & 0.500 & 0.595 & 0.272 & 0.419 \\
rock \textbf{v.s.} classical  & 0.478 & 0.667 & 0.683 & 0.533 & 0.627 & 0.520 \\
hiphop \textbf{v.s.} jazz     & 0.583 & 0.500 & 0.720 & 0.627 & 0.391 & 0.482 \\
hiphop \textbf{v.s.} classical& 0.429 & 0.560 & 0.686 & 0.571 & 0.557 & 0.548 \\
jazz \textbf{v.s.} classical  & 0.436 & 0.600 & 0.700 & 0.295 & 0.829 & 0.384 \\
\bottomrule
\end{tabular}
\caption{Cross-genre preference comparison results. Pred denotes the model-predicted win rate of the first genre in each pair, GT denotes the ground-truth human win rate, and Acc. denotes pairwise accuracy. The results reveal systematic and directional discrepancies, including consistent overestimation of pop and underestimation of genres such as jazz.}
\label{tab:cross_genre}
\end{table*}

\subsection{Aesthetic Scores Align with Pop-Centered Directions}
\label{subsec:aesthetic}
The analyses above consistently reveal a pop-centered pattern: pop and stylistically related genres tend to receive higher scores, while genres such as jazz and classical are often disadvantaged. To examine this phenomenon more rigorously, we test whether the model predictions are aligned with a pop-related direction in the learned feature space.

Specifically, we train a lightweight binary classifier to distinguish pop from non-pop samples, and use its logits as a measure of pop proximity. Across both datasets, pop proximity shows a strong monotonic relationship with predicted aesthetic scores. The Spearman rank correlation reaches 0.88 on M6 and 0.81 on MTG-Jamendo, which indicates that model predictions are tightly aligned with a pop-related direction.

\section{Proposed Method}
\label{sec:method}

Section~\ref{sec:diagnostic} shows that SongEval exhibits obvious shortcut learning behavior. Instead of relying purely on intrinsic acoustic cues relevant to musical aesthetics, the model tends to exploit correlations between genre-related features and score. As a result, samples associated with dominant patterns become easier to predict, while others remain under-optimized, leading to systematic imbalance in prediction quality across the data distribution.

To mitigate this issue, we design a training objective that explicitly counteracts this optimization bias. The key observation is that shortcut learning manifests in how the model allocates its learning capacity: easy samples dominate the optimization, and groups aligned with shortcut directions accumulate lower errors over time. We therefore regulate the training process at both the sample and group levels in a unified manner.

Formally, given a dataset $\mathcal{D}=\{(x_i, y_i, g_i)\}_{i=1}^{N}$, where $x_i$ denotes an audio sample, $y_i \in \mathbb{R}^d$ is the ground-truth aesthetic score vector, and $g_i \in \mathcal{G}$ is the genre label, the model predicts $\hat{y}_i = f_\theta(x_i)$.

For each sample, we define its regression error as
\begin{equation}
e_i = \frac{1}{d}\|\hat{y}_i - y_i\|^2.
\end{equation}

Under the standard MSE objective, optimization is dominated by samples with small errors, which are typically those aligned with shortcut patterns. To counter this effect, we adopt a focal-style regression objective:
\begin{equation}
\mathcal{L}_{\text{focal}} =
\frac{1}{N}\sum_{i=1}^{N} e_i^{\gamma + 1},
\end{equation}
where $\gamma \ge 0$ controls the focusing strength. This formulation amplifies the contribution of high-error samples, forcing the model to move beyond easily exploitable correlations and better fit harder regions of the data distribution.

To further characterize prediction quality in a form that is comparable across groups, we define a differentiable performance surrogate:
\begin{equation}
s_i = -e_i,
\end{equation}
where higher values indicate better prediction performance.

For each group $k \in \mathcal{G}$, we compute the average performance within a mini-batch:
\begin{equation}
g_k = \frac{1}{|\mathcal{B}_k|} \sum_{i \in \mathcal{B}_k} s_i,
\end{equation}
where $\mathcal{B}_k$ denotes the set of samples in group $k$ within the current batch.

However, training on full-length music recordings severely limits the batch size due to memory constraints, making $g_k$ highly noisy and unreliable. To obtain stable estimates of group-level performance, we maintain exponential moving averages (EMA) of both global and group-wise statistics:
\begin{equation}
\mu \leftarrow \beta \mu + (1 - \beta)\,\bar{s}, \quad
\mu_k \leftarrow \beta \mu_k + (1 - \beta)\, g_k,
\end{equation}
where $\bar{s}$ is the batch-average performance and $\beta$ is the decay factor.

Based on these historical estimates, we measure the deviation of each group from the global performance:
\begin{equation}
\Delta_k = \mu_k - \mu.
\end{equation}

This deviation reflects how the model distributes its optimization effort across groups. In the presence of shortcut learning, groups aligned with dominant patterns tend to achieve consistently better performance (higher $\mu_k$), while others lag behind.

While the focal loss increases attention to hard samples, it does not explicitly prevent such group-level imbalance from persisting. Therefore, we further introduce a group-level constraint to regulate how performance is distributed across groups.

Specifically, we assign adaptive weights based on historical deviation:
\begin{equation}
w_k = 1 + |\Delta_k|,
\end{equation}
and define the group-level regularization as:
\begin{equation}
\mathcal{L}_{\text{group}} = \frac{1}{|\mathcal{G}|} \sum_{k \in \mathcal{G}} w_k \cdot ( g_k - \mu )^2.
\end{equation}

This objective penalizes discrepancies between group-wise performance and the global reference, encouraging the model to allocate learning capacity more evenly across the data distribution instead of over-optimizing groups that are already well aligned with shortcut directions.

Finally, the overall training objective is given by:
\begin{equation}
\mathcal{L} = \mathcal{L}_{\text{focal}} + \lambda_{\text{group}} \cdot \mathcal{L}_{\text{group}},
\end{equation}
where $\lambda_{\text{group}}$ controls the strength of the group-level regularization.

In summary, the proposed objective counteracts shortcut learning through a unified mechanism: the focal formulation reshapes the sample-wise gradient distribution to emphasize hard examples, while the group-level constraint regulates the allocation of optimization effort across groups. Together, they reduce the model's reliance on shortcut correlations and promote more balanced and robust learning over the full data distribution.

\begin{table*}[t]
\centering
\small
\setlength{\tabcolsep}{20pt}
\begin{tabular}{lcccccc}
\hline
& \multicolumn{3}{c}{CMI-Pref (Ours)} & \multicolumn{3}{c}{CMI-Pref-Pseudo (Ours)} \\
Pair & Pred & GT & Acc. & Pred & GT & Acc. \\
\hline
pop \textbf{v.s.} rock & 0.806 & 0.645 & 0.635 & 0.550 & 0.245 & 0.575 \\
pop \textbf{v.s.} hiphop & 0.798 & 0.832 & 0.826 & 0.730 & 0.594 & 0.724 \\
pop \textbf{v.s.} jazz & 0.462 & 0.462 & 0.848 & 0.761 & 0.249 & 0.399 \\
pop \textbf{v.s.} classical & 0.843 & 0.675 & 0.696 & 0.794 & 0.412 & 0.523 \\
rock \textbf{v.s.} hiphop & 0.679 & 0.893 & 0.712 & 0.469 & 0.660 & 0.609 \\
rock \textbf{v.s.} jazz & 0.000 & 0.250 & 0.500 & 0.590 & 0.272 & 0.468 \\
rock \textbf{v.s.} classical & 0.478 & 0.667 & 0.732 & 0.583 & 0.627 & 0.561 \\
hiphop \textbf{v.s.} jazz & 0.583 & 0.500 & 0.740 & 0.606 & 0.391 & 0.496 \\
hiphop \textbf{v.s.} classical & 0.440 & 0.560 & 0.730 & 0.600 & 0.557 & 0.575 \\
jazz \textbf{v.s.} classical & 0.436 & 0.600 & 0.691 & 0.350 & 0.829 & 0.404 \\
\hline
\end{tabular}
\caption{Cross-genre results of our proposed method.}
\label{tab:cross_genre_ours}
\end{table*}

\section{Experiments}
\label{sec:experiments}

\subsection{Experimental Setup}

We evaluate the proposed method under the same data and evaluation protocol described in Section~\ref{sec:diagnostic}. Following the standard SongEval setup, the baseline model is trained using the MSE objective with the Adam optimizer, a learning rate of $3\times10^{-5}$, for 30 epochs. Due to the use of full-length audio inputs, the batch size is set to 1, and all experiments are conducted on eight NVIDIA H20 GPUs.

We then retrain the model using the training objective introduced in Section~\ref{sec:method}, which augments the standard regression loss with a focal formulation and a group-level regularization term to counteract shortcut-driven optimization bias. The focal loss parameter is set to $\gamma = 2$, and the group-level regularization coefficient is set to $\lambda_{\text{group}} = 1 \times 10^{6}$. To stabilize optimization, we apply a linear warm-up strategy to the group-level regularization coefficient $\lambda_{\text{group}}$, increasing it from 0 to its target value over the first 5 epochs and keeping it constant thereafter.

\subsection{Evaluation Metrics}

We evaluate the model from three complementary perspectives, progressing from optimization dynamics to behavioral outcomes and finally to structural biases.

We first analyze the training dynamics to understand how different objectives affect the optimization process. Specifically, we track the EMA performance statistics $\mu_k$ for each genre throughout training.

We then evaluate how these changes in optimization dynamics translate into prediction behavior. Following the preference-based evaluation protocol in Section~\ref{subsec:preference}, we measure alignment with human preferences using pairwise comparisons from the CMI dataset, reporting results for both same-genre and cross-genre pairs.

Finally, we examine whether the model's predictions remain coupled with genre-related signals at a global level. Specifically, we compute the average score for each genre to derive a genre-level ranking, and measure the SRCC between this ranking and a reference ranking derived from external signals.

\begin{figure}[t]
\centering
\includegraphics[width=\linewidth]{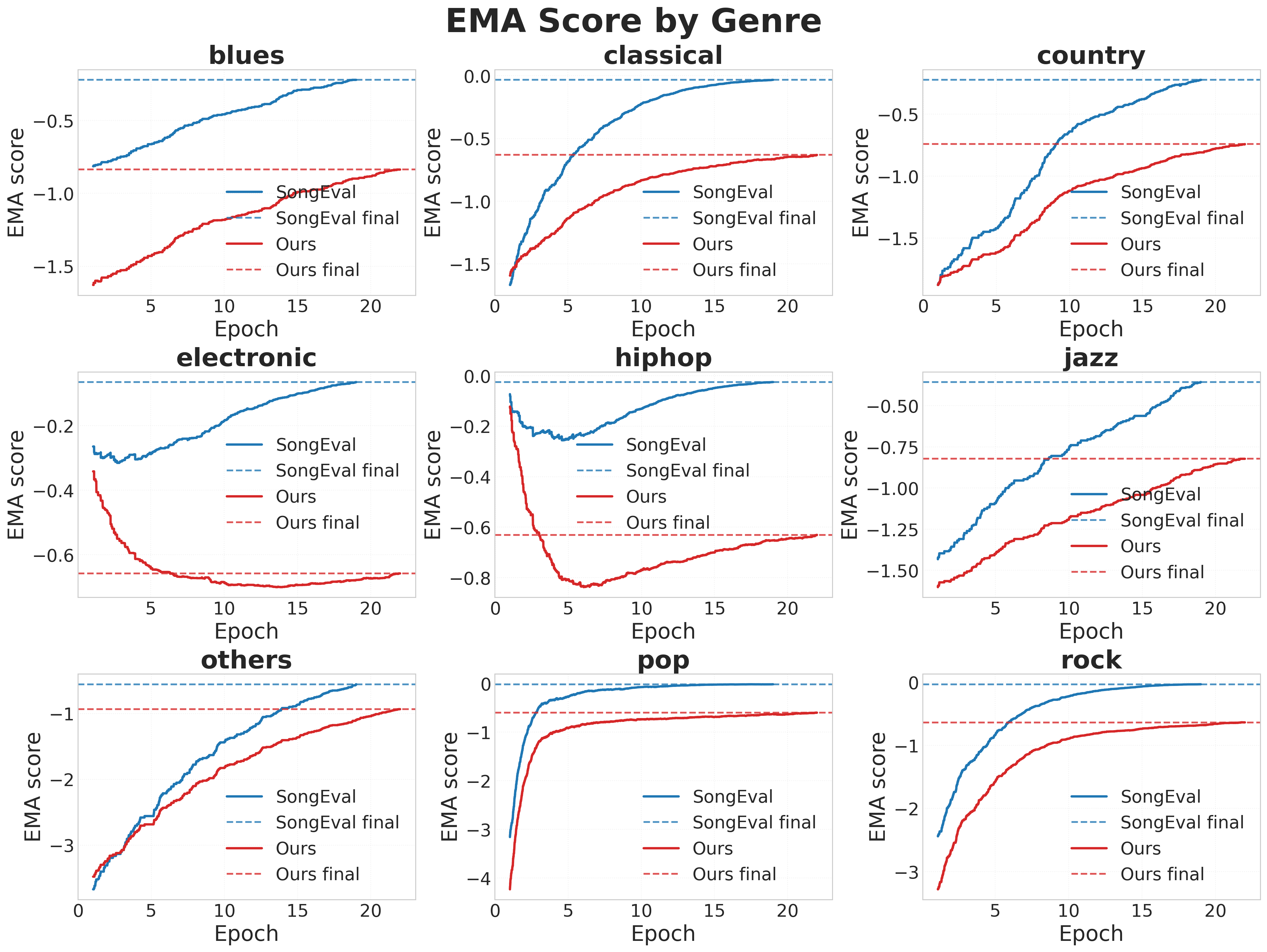}
\caption{EMA score trajectories across genres during training. The baseline (top) exhibits a shortcut-driven dynamic, where dominant genres (e.g., pop) improve rapidly while others (e.g., jazz and country) lag behind. In contrast, our method (bottom) leads to more synchronized and balanced improvement across genres.}
\label{fig:ema_genre}
\end{figure}

\subsection{Results}

We first examine the optimization dynamics during training, as illustrated in Figure~\ref{fig:ema_genre}, in the baseline, training is dominated by shortcut-driven dynamics, where dominant genres such as pop improve rapidly in early stages, while other genres, including jazz and country, improve much more slowly, resulting in persistent performance gaps throughout training, in contrast, under the proposed objective, the EMA trajectories become significantly more synchronized across genres, different genres progress at a more consistent rate, and the gap between fast- and slow-learning genres is substantially reduced, indicating that the model no longer prioritizes a subset of dominant patterns but instead learns more uniformly across genres.

We further evaluate the impact of this change on preference prediction. Compared to the baseline, the overall accuracy improves from 0.687 to 0.715 on CMI-Pref and from 0.549 to 0.566 on CMI-Pref-Pseudo. Consistent gains are observed in both same-genre comparisons, from 0.669 to 0.703 on CMI-Pref and from 0.563 to 0.579 on CMI-Pref-Pseudo, and cross-genre comparisons, from 0.713 to 0.735 on CMI-Pref and from 0.527 to 0.545 on CMI-Pref-Pseudo, indicating that the improvement extends beyond within-genre comparisons and generalizes to heterogeneous cases.
These improvements align with the structured discrepancies identified in Section~\ref{sec:diagnostic}. In particular, cross-genre comparisons become more stable. As shown in Table~3, performance on challenging pairs, especially those involving jazz and classical, improves consistently.
In addition, the directional bias observed in the baseline is partially mitigated. As discussed earlier, the baseline systematically overestimates pop and underestimates jazz. Under the proposed objective, these discrepancies are reduced in several cases. For instance, in pop--jazz comparisons on CMI-Pref-Pseudo, the gap between predicted and ground-truth win rates decreases from 0.795 / 0.249 to 0.761 / 0.249, indicating a reduction in pop overestimation. Similarly, the underestimation of jazz in jazz--classical comparisons is alleviated, from 0.315 / 0.829 to 0.350 / 0.829.

At the same time, well-aligned pairs such as pop--hiphop remain stable, suggesting that the improvement does not come at the cost of degrading already reliable comparisons.

Additionally, the strong alignment between model predictions and the pop-related direction is substantially reduced. Specifically, the Spearman rank correlation decreases from 0.88 to 0.68 on M6 and from 0.81 to 0.62 on MTG-Jamendo. This indicates that the model no longer relies as heavily on the pop-related direction, further suggesting that the dependence on genre-related shortcuts has been effectively mitigated.

Finally, we provide a brief ablation study. Due to space constraints, we report only the overall results.

When using only the focal objective, the model achieves moderate improvements over the baseline, with overall accuracy reaching 0.699 on CMI-Pref and 0.556 on CMI-Pref-Pseudo. Gains are observed in both same-genre comparisons, where accuracy reaches 0.679 on CMI-Pref and 0.569 on CMI-Pref-Pseudo, and cross-genre comparisons, where accuracy reaches 0.729 on CMI-Pref and 0.534 on CMI-Pref-Pseudo.

When using only the regularization term, the improvement becomes stronger, with overall accuracy reaching 0.705 on CMI-Pref and 0.564 on CMI-Pref-Pseudo. Performance also improves in same-genre comparisons, reaching 0.686 on CMI-Pref and 0.577 on CMI-Pref-Pseudo, and in cross-genre comparisons, reaching 0.733 on CMI-Pref and 0.544 on CMI-Pref-Pseudo.

Overall, these results show that the proposed objective not only improves predictive accuracy, but also leads to more balanced and consistent preference modeling across genres, partially mitigating the directional bias and inconsistency identified in earlier analysis, and these improvements are closely linked to the more uniform optimization dynamics observed during training.

\section{Conclusion}

We identify a critical failure mode in music aesthetics evaluation: genre-induced shortcut learning. Through a progressive diagnostic analyses, we show that models rely on genre-related signals as proxies for musical aesthetics, further leading to pop-centric bias.

We further demonstrate that this behavior is propagated in shortcut-driven optimization dynamics, where dominant patterns are preferentially learned and reinforced. To mitigate this issue, we propose a unified objective that combines focal reweighting with group-level regularization, encouraging more balanced learning across genres.

Experimental results demonstrate that the proposed objective reduces genre-dependent bias and improves alignment with human preferences, with consistent gains in both within-genre and cross-genre preference comparisons. 

\section{Limitations}

First, model performance is fundamentally constrained by data. While our method mitigates shortcut learning during optimization, a more diverse dataset with reliable and balanced annotations remains essential.

Second, beyond genre, models may still exploit other spurious cues (e.g., loudness or instrumentation). Identifying and characterizing such hidden biases remains an important direction for future work.

\bibliography{ISMIRtemplate}

%
%
%
%

\end{document}